\documentclass[twocolumn,aps,prb,showpacs,raggedbottom,nobalancelastpage,amssymb]{revtex4}
\usepackage{amsmath}
\usepackage{amsfonts}
\usepackage{graphicx}
\usepackage{appendix}
\usepackage{dsfont}
\usepackage{amssymb}
\usepackage{bm}
\usepackage{color}
\usepackage{ulem}
\usepackage{soul}
\usepackage{natbib}
\setstcolor{red}
\usepackage{nccmath}
\usepackage{array}

\begin{document}

\title{Spin-orbit coupling and Anomalous Josephson effect in Nanowires}

\author{G. Campagnano$^{1,2}$}
\author{P. Lucignano$^{2,1}$}
\author{D.Giuliano$^{3,2}$}
\author{A. Tagliacozzo$^{1,2,4}$}
\begin{abstract}
A superconductor-semiconducting nanowire-superconductor heterostructure in the presence of spin orbit coupling and magnetic field can  support a  supercurrent even in the absence of phase difference between the superconducting electrodes. We investigate this phenomenon---the  anomalous Josephson effect---employing a model capable of describing many bands in the normal region. We discuss geometrical and symmetry conditions required to have finite anomalous supercurrent and in particular  we show that this phenomenon is enhanced when the Fermi level is located close to a band opening in the normal region.
\end{abstract}

\affiliation{$^1$ Dipartimento di Scienze Fisiche, Universit\`a di Napoli ``Federico II'',
Monte S.Angelo, I-80126 Napoli, Italy}
\affiliation{$^2$ CNR-SPIN, Monte S.Angelo  via Cinthia,  I-80126 Napoli, Italy}
\affiliation{$^3$ Dipartimento di Fisica, Universit\`a della Calabria and INFN, Gruppo 
Collegato di Cosenza Arcavacata di Rende, I-87036, Cosenza, Italy} 	
\affiliation{$^4$ INFN, Laboratori Nazionali di Frascati, Via E. Fermi 40, 00044 Frascati, Italy}

\date{\today}

\maketitle

\section{Introduction}

Hybrid nanostructures, combining superconducting and semiconducting  elements,  hold promise for novel functionalities and their transport properties are attracting increased attention and intense experimental and theoretical effort. A two-dimensional  electron gas trapped in a semiconductor  heterostructure,  and contacted with superconductors, provides an archetypical  Superconductor/ Normal metal/ Superconductor (SNS)  system\cite{Takayanagi_PRL_1995}. Semiconducting quantum dots \cite{DeFranceschi:2010} or semiconducting nanowires \cite{Doh:2005,Sand-Jespersen:2007,vanDam:2006} can be contacted with superconductors.  These devices allow for gating of the semiconductor with increased  control of its carrier density\cite{Morpurgo:2012}.  In this respect, hybrid structures including ferromagnetic barriers  offer  another exciting arena on its own\cite{BuzdinRMP, GolubovRMP}  and are  of the utmost relevance for Spintronics\cite{Awschalom:2013,Crepaldi:2012}
Recently, there has been a burst of activity on  hybrid  heterostructures involving  semiconductors  with strong spin-orbit (SO) interaction, as SO can be  a tremendous tool for controlling spin transport, as well. It  has been established that SO  is at the origin of  a topological non trivial order in
 Topological Insulator    materials respecting time reversal symmetry  \cite{Bernevig:2006,Konig:2007,Kane:2007,Hsieh:2008}. Semiconducting nanowires, with SO interaction, Zeeman spin splitting induced by a longitudinal magnetic field, and superconductivity, can be described in terms of a chain of spinless fermions. It has been shown that zero energy neutral excitations (Majorana Bound States) localize  at the end of the nanowire, when the system is driven to the topological non trivial state\cite{Kitaev:2001,Lutchyn:2010,Oreg:2010,Kouwenhoven:2012}.  As Majorana Bound States can become a valuable tool in future Quantum Information devices \cite{Nayak:2008},  it is becoming more and more essential nowadays  to characterize the superconducting proximity in  quasi one-dimensional SNS structures with  SO and Zeeman spin splitting both in the topologically trivial phase and in the non trivial one.

Here we consider a semiconductor nanowire with both Rashba and Dresselhaus  SO coupling, forming a quasi one-dimensional wire, contacted with   conventional  s-wave  singlet pairing  superconductors.  We assume that the normal region is much shorter than the superconducting coherence length $\xi$ (short junction limit),  so that  the superconducting coherence is fully established across the SNS structure. In the short junction limit,  the Josephson current is carried by the Andreev bound states, belonging to the discrete  subgap  energy spectrum \cite{Beenakker:1991}.

The anomalous Josephson Effect (AJE), consists of a non zero Josephson current  $I_a$ flowing with zero phase difference between the superconducting order parameters of the contact superconductors, $\varphi=0$.  This implies that the zero current ground state is located at a phase difference $\varphi_0 \neq \{0,\pi\}$ (in some literature these junctions are named $\varphi_0$-junctions\cite{Buzdin:2008}).
The situation
considered here is different from the case of Josephson junctions with a strong negative second 
harmonic\cite{Goldobin:2007}. For these junctions the ground state is located at a  phase difference determined by the minimum of the Josephson energy, which  depends on the relative strength of the first and the second harmonic.

 The AJE was initially predicted for unconventional 
superconductors\cite{Larkin:1986, Satoshi:2000, Sigrist:1998,PhysRevB.52.3087, PhysRevLett.102.227005, PhysRevLett.99.037005}.
Several possibilities for the normal region have been addressed: a magnetic normal metal\cite{Buzdin:2008,PhysRevB.67.184505,Nat_phys_Eschrig_2008,PhysRevB.76.224525,PhysRevLett.98.077003,PhysRevLett.102.017001}, a one-dimensional quantum wire, a quantum dot\cite{Zazunov:2009,brunetti:2013}, a multichannel system with a barrier or a quantum point contact\cite{Reynoso:2008,Reynoso:2012}, a semiconducting nanowire\cite{Yokoyama:2013,Yokoyama:2014}. Anomalies of the Josephson current have also been predicted  in presence of Coulomb interactions  and SOI  for a wire \cite{Krive:2004,Krive:2005}  or a Quantum Dot \cite{brunetti:2013} contacted with conventional s-wave superconductors. 

We show that  $I_a$ is maximum when the system's parameters are away from highly symmetrical points ( i.e. pure Rashba, pure Dresselhaus or equal strength of Rashba and Dresselhaus SO interaction). Moreover, while  studying   the dependence of  $I_a$  on the location of the chemical potential $E_F$, we find  an enhancement of  $I_a$ when $E_F$ is close to the opening of a new channel in the normal region. Similar threshold effects are not new in scattering theory, even in the absence of spin effects\cite{Tagliacozzo:2009}.  

The paper is organized as follows: in Section II we review the relation between the Josephson current and the scattering
matrix of the normal region. In Section III we show how to compute the scattering matrix as a function of the chemical potential, the spin-orbit 
coupling and the Zeeman field. In Section IV we discuss our results, we summarize in Section V. 

\section{Andreev states and current calculation}
 \begin{figure}[htbp]
\begin{center}
\includegraphics[width=0.9\linewidth]{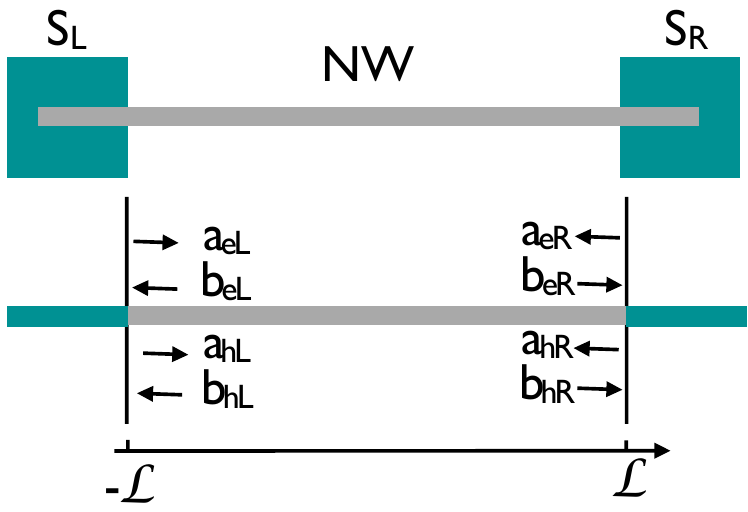}
\caption{Electron and hole incoming states in the ($ a_i$) are scattered into outgoing states ($ b_i$) 
by nanowire region described by the scattering matrix $S_{e/h}$. 
Andreev scattering at NS interfaces where electron (hole) states are scattered into hole (electron) 
states is described by the Andreev scattering matrix $S_A$.}
\label{fig1}
\end{center}
\end{figure}

We consider a quasi 1D nanowire between two conventional s-wave superconductors. In our model 
the nanowire is in the region $|x|<\mathcal{L}$,
while the left (right) superconductor is in the region $x<-\mathcal{L}$ ($x>\mathcal{L}$). We assume 
that the SO interaction and the 
magnetic field are only present in the nanowire region. 
Recent advances in the fabrication of nanowires of InAs and InSb have made 
materials with a large g-factor and a strong spin-orbit (SO) interaction available.
A topological phase with Majorana fermions as boundary excitations could appear when the SO interaction was present in the leads\cite{Alicea:2012}. 
By contrast, we assume no SO interaction in the leads as we want to focus on the non-topological phase, which has been poorly addressed in the literature.
Moreover due to a large $g$ factor in the
semiconducting region, a strong Zeeman spin splitting can be recovered, even for weak values of the magnetic field
which, in turn, does not affect the superconductors and is accordingly  neglected in the leads. Moreover we disregard the 
orbital effect of the magnetic field, which we take perpendicular to the plane containing the nanowire and the top surface of the leads.

We look for  solutions of the Bogoliubov-de Gennes equations:
\begin{multline}
\label{bdg}
\left(
\begin{array}{cc}
H-E_F & \Delta \\
 \Delta^\dag & -(H^*-E_F)
\end{array}
\right)
\left(
\begin{array}{c}
u(x,y) \\ v(x,y)
\end{array}
\right)
= \epsilon 
\left(
\begin{array}{c}
u(x,y) \\ v(x,y)
\end{array}
\right),
\end{multline}
where $\epsilon$ measures the energy with respect to the Fermi level $E_F$, while $u(x,y)$ and $v(x,y)$ are
respectively the electron and hole spinors in the Nambu representation.
Notice that the spin structure is not explicit at this level. The even parity pairing potential is  
taken to be: 
\begin{equation}
\hat{\Delta}=\Delta(x)\left(
\begin{array}{cc}
0 & -1 \\
1 & 0
\end{array}
\right)\:,
\end{equation}
with 
\begin{equation}
\Delta(x)=\Delta_0 \left[  \Theta(-x-\mathcal{L})e^{-i \varphi/2}+\Theta(x-\mathcal{L})e^{i \varphi/2} \right],
\end{equation}
and we take a symmetric phase difference $\varphi$ between the two superconductors. $\Theta (x)$
is the Heaviside step function. 

For the sake of generality we assume that both Rashba and Dresselhaus SO interaction are present  \cite{note}.
We notice that in conventional InAs/InSb nanowires the Rashba term is in general  much larger than the Dresselhaus one.
However, in our calculation we also explore a region of parameters where the two 
interactions have comparable strength, which is achievable in (In,Ga)As quantum wells \cite{Kohda_prb:2012}. We assume an harmonic confining potential in the $y$ directions, so that 
electrons propagate along the  the $x$ direction . The choice of an harmonic confining potential  is particularly 
convenient when expressing the matrix elements of the SO interaction.\cite{Governale:2002,marigliano:2004,nota2} The Hamiltonian reads:

\begin{multline}\label{hamiltonian}
H=\frac{p_x^2}{2m}+\frac{p_y^2}{2m}+\frac{1}{2}m \omega^2 y^2 +\frac{\alpha(x)}{\hbar}(\sigma_x p_y-\sigma_y p_x)
\\
+\frac{\beta(x)}{\hbar}(\sigma_x p_x-\sigma_y p_y)+g \mu_B B \sigma_z+\delta_a\delta(x+\mathcal{L})
\\ + \delta_b\delta(x-\mathcal{L})+
\frac{i}{2}(\partial_x \alpha(x) \sigma_y-\partial_y \beta(x) \sigma_x),
\end{multline}
where $\alpha(x)=\alpha[\Theta(x+\mathcal{L})-\Theta(x-\mathcal{L})]$ and $\beta(x)=\beta[\Theta(x+\mathcal{L})-
\Theta(x-\mathcal{L})]$
are respectively the strength of the Rashba and Dresselhaus component of the SO interaction. An important
ingredient in Eq.~\ref{hamiltonian} is given by the $\delta$-like scattering centers, with 
strength $\delta_a$ and $\delta_b$ different from each other. This point is crucial, here, as 
such an asymmetry between left and right contact is fundamental in determining the anomalous Josephson effect, 
as we discuss in the following. Finally, $B$ is the applied magnetic field, while 
the last term in Eq.~\ref{hamiltonian} is necessary to preserve the hermicity of the Hamiltonian 
operator  in case of a position-dependent SO interaction strength. 
Due to the  presence of the transverse confinement potential, it is natural to employ 
a scattering approach to the problem.  Indeed in the superconducting leads 
one can define transverse modes, so that the electron wave functions in the $x$-direction
are  characterised by a band and a spin index. 

The spectrum of Eq.~\ref{bdg} consists of a finite set of bound states (Andreev levels) with energy $|\epsilon|<\Delta_0$,
and a continuum of states with $|\epsilon|>\Delta_0$. The current can be obtained from the ground 
state energy $E_{gs}(\varphi)$ at zero temperature by the thermodynamic relation 
\begin{equation}
I(\varphi)=\frac{2e}{\hbar}\frac{d E_{gs}}{d\varphi}.
\end{equation}
In the short junction limit only the 
subgap Andreev states contribute to the Josephson current (in the long-junction limit, one can use the technique developed in 
Refs.[\onlinecite{giuliano:2013,giuliano:2014}] to exactly account for contributions from all the states, to leading order in the inverse junction length). Thus, one obtains   
\begin{equation}
I(\varphi)=\frac{e}{\hbar}\sideset{}{'}\sum_n  \frac{\partial E_n(\varphi)}{\partial \varphi}.
\label{cur1}
\end{equation}
In Eq.~\ref{cur1}  "$n$" labels the Andreev states, the primed sum means that only negative energy (occupied) Andreev states 
are  considered. Notice that a factor of 2 difference with the usual 
relation found in literature because  spin degeneracy is lifted here.
Before we move further in our analysis it is useful to notice, as put forward in
Refs.[\onlinecite{Liu_symmetries_PRB:2010,Yokoyama:2013}], how the symmetry of Eqs.~\ref{bdg}
can rule out the anomalous Josephson effect.
Denoting with $\mathcal{H}_{BdG}$ the matrix in Eq.~\ref{bdg}, in the absence of magnetic field,  the time reversal 
operator $\mathcal{T}=i \sigma_y K$ (with $K$ the complex conjugation operator) induces a self-duality 
$\mathcal{T}\mathcal{H}_{BdG}(\varphi)\mathcal{T}^{-1}=\mathcal{H}_{BdG}(-\varphi)$, so 
that, if $\mathcal{H}_{BdG}(\varphi)$ has an  allowed energy eigenvalue  $E_n$, 
$\mathcal{H}_{BdG}(-\varphi)$ will have the same eigenvalue. In this case 
the anomalous Josephson effect is ruled out. Similarly, when there is no  
SO interaction, one obtains $K\mathcal{H}_{BdG}(\varphi)K^{-1}=\mathcal{H}_{BdG}(-\varphi)$ 
which again implies no anomalous Josephson effect. The previous considerations are general, not depending 
on the details of the Hamiltonian describing the nanowire (explicit form  of the confinement
potential, presence of disorder, etc.).
We notice that for our specific model, even in presence of SO interaction and magnetic field along the $z$ 
direction, when $\delta_a=\delta_b$ the operator $\mathcal{O}=i \sigma_z \Pi$ (with $\Pi$ the parity operator) 
also gives $\mathcal{O}\mathcal{H}_{BdG}(\varphi)\mathcal{O}^{-1}=\mathcal{H}_{BdG}(-\varphi)$; 
for this reason, in order to recover the anomalous Josephson effect,  in this Article we 
always assume asymmetric delta potentials at the
interfaces between normal and superconducting regions. 
 
The calculation of the Andreev states in a quasi one dimensional system can be conveniently done
using a scattering matrix approach \cite{Beenakker:1991}. We assume that 
for energies below the superconducting gap $\Delta_0$
at the interface between the normal and the superconducting regions only
intra-channel Andreev scattering takes place 
where a hole  (electron) with spin $\sigma$
is reflected as an electron (hole) with spin $-\sigma$.
Accordingly, these processes are encoded in the 
relations

\begin{equation}
\left(\begin{array}{c}
a_{eL} \\
a_{eR} \\
a_{hL} \\
a_{hR}
\end{array}
\right) =S_A
\left(\begin{array}{c}
b_{eL} \\
b_{eR} \\
b_{hL} \\
b_{hR}
\end{array}
\right)
\end{equation}
and the Andreev scattering matrix is defined as:
\begin{equation}
S_A = 
\left(\begin{array}{cc}
0 &  \hat r_{eh}\\
\hat r_{he}    & 0  
\end{array}
\right)
\end{equation}
with 
\begin{equation}
\hat r_{eh} =  i\,
e^{-i \gamma }\left(\begin{array}{cc}
 \hat 1 \otimes \hat \sigma_y e^{- i \varphi/2}&  0\\
0    &  \hat 1 \otimes \hat \sigma_y e^{+ i\varphi/2}
\end{array}
\right)\, ,
\end{equation}
and
\begin{equation}
\hat r_{he} = - i\,
e^{-i \gamma }\left(\begin{array}{cc}
 \hat 1 \otimes \hat \sigma_y e^{+i \varphi/2}&  0\\
0    &  \hat 1 \otimes \hat \sigma_y e^{- i\varphi/2}
\end{array}
\right)\, .
\label{ll.1}
\end{equation}
In Eq.~\ref{ll.1}  $\hat 1$ is the identity matrix in the channel space, 
the $\hat \sigma_y$ Pauli matrix acts in 
the spin space and $\gamma=\arccos(\epsilon/\Delta)$.  In the normal
region Andreev scattering processes are not allowed, which 
allows us to write
\begin{equation}
\left(\begin{array}{c}
b_{eL} \\
b_{eR} \\
b_{hL} \\
b_{hR}
\end{array}
\right)
=\left(
\begin{array}{cc}
S_e(\epsilon) & {\bf 0} \\
{\bf 0} & S_h(\epsilon)  
\end{array}
 \right) 
 \left(\begin{array}{c}
a_{eL} \\
a_{eR} \\
a_{hL} \\
a_{hR}
\end{array}
\right).
 \end{equation} 
The energy of the Andreev bound states is determined by the 
following equation \cite{Beenakker_universal}
\begin{equation}\label{determinant}
\det \left[ \hat{1} -\hat{r}_{eh} \hat{S}_{h}(-\epsilon)  \hat{r}_{he}   \hat{S}_{e}(\epsilon) \right]=0.
\end{equation}
In the short-junction limit one can disregard the energy dependence of the scattering matrix 
and take $ \hat{S}^*_{h}(-\epsilon)=\hat{S}_{e}(\epsilon)\simeq\hat{S}_{e}(0) $. 
Therefore in order to solve Eq.~\ref{determinant}, one has to calculate 
the scattering matrix of the normal region at the Fermi energy. We carry on this task  in the next Section.

\begin{figure}
\includegraphics[width=0.9\linewidth]{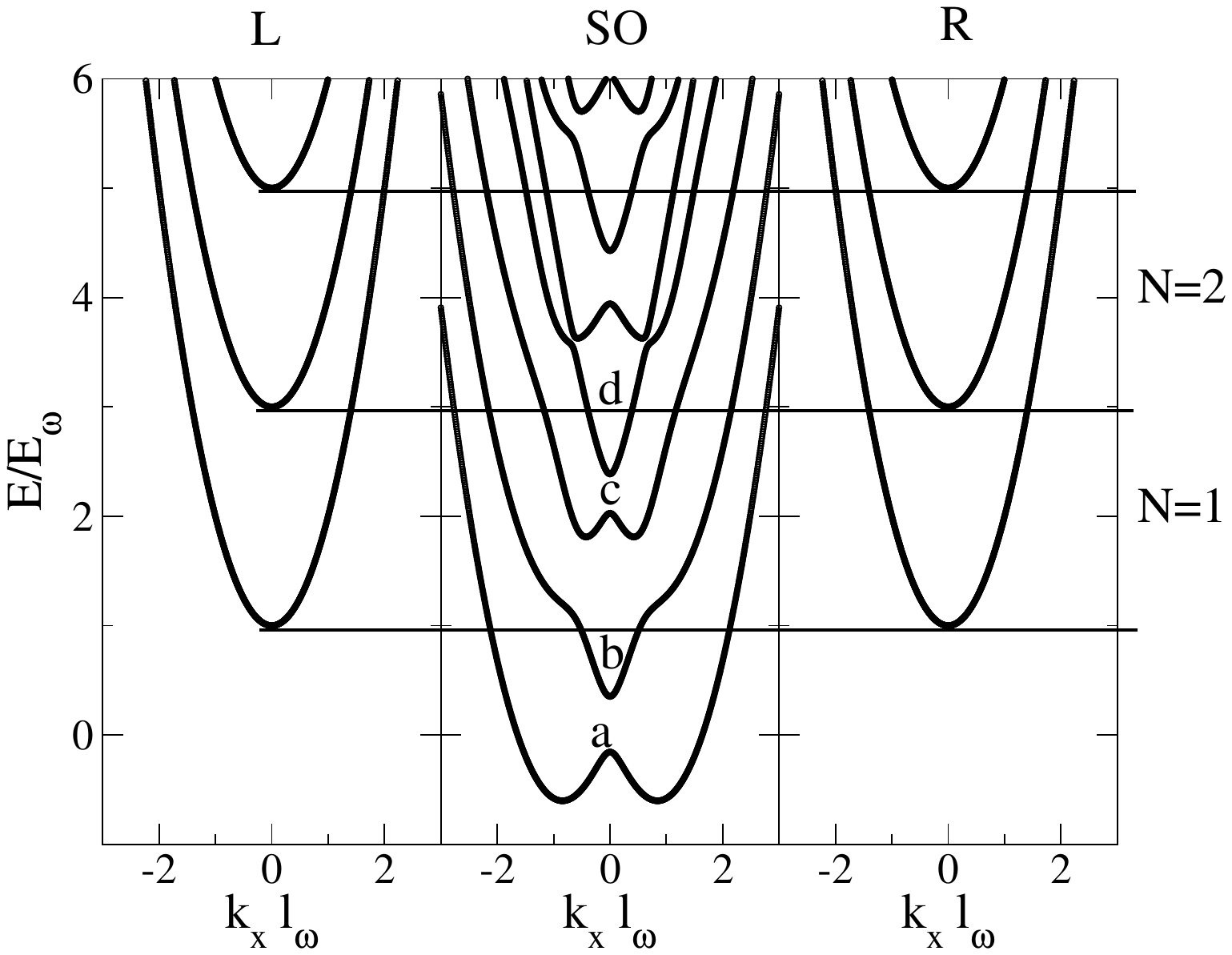}
\caption{Here we show the band structure obtained by numerically diagonalising Hamiltonian of 
Eq.~\ref{hamiltonian}, we define $E_\omega=\hbar \omega/2$ and $l_\omega=\sqrt{\hbar/m\omega}$. 
In this figure we use the following paramenters: $m \zeta l_\omega/\hbar^2=0.9$, $\theta=\pi/8$, 
$g \mu_B B/E_\omega=0.6$ and the number of transverse bands $l=5$.}
\label{nw-band-structure}
\end{figure}

\section{Model and calculation of the scattering matrix S}

In this Section we evaluate the scattering matrix S of the normal region as  a function of the
Fermi energy. 
We address the case of one open channel (with two spin orientations). However, 
our method is easily generalised 
to two or more open channels. 

As a first step, we solve the Schroedinger equation within 
the L-lead ($x<-\mathcal{L}$)  and the R-lead ($x>\mathcal{L}$) (assuming no superconductivity), as well as
in the central region $SO$ with spin-orbit interaction ($x<|\mathcal{L}|$)  
(see Fig.s \ref{fig1},\ref{nw-band-structure}). Therefore, we  
derive  the scattering matrix by matching the solutions  at the interfaces between the 
three  regions. In Fig.~\ref{nw-band-structure} we plot the band structure in these three regions  (parameters in the caption).  One can notice that the band structure is strongly influenced by the presence of the spin orbit coupling and, interestingly enough, depending on the position of the Fermi level, the number of open channels  of the leads and the central region can be different

To obtain a reliable approximation for  the eigenfunctions and eigenvectors in 
the SO region, we exactly diagonalize
the Hamiltonian in the basis of the wave functions 
$\{e^{i \kappa x} \chi_m(y)\phi_\sigma\}$, with $\{\chi_m(y)\}$ being the 
eigenfunctions of the harmonic oscillator  and $\phi_\sigma  $ the eigenfunctions of $\sigma_z$, by considering  
a finite number $l$ of transverse modes $\{\chi_m(y)\}$, $m\in\{1,...,l\}$.
The above basis functions are eigenfunctions in the right (left) region R (L). 

Such an approximation is indeed expected to provide a 
good description of the scattering dynamics of the system at  energies small 
compared to the energy of the last   transverse 
mode considered. In our calculation we 
find that  the low energy properties of our system are well described if we truncate the basis at  $l = 5$. 

Therefore, the problem is now reduced to finding the eigenvalues 
and the eigenfunction of the corresponding $2l \times 2l$-dimensional 
Hamiltonian matrix
$\mathcal{H}(\kappa)_{m,\sigma;m',\sigma'}$. For a given energy E 
the allowed $\kappa_i$ are the solution of
\begin{equation}
det \left[ \mathcal{H}(\kappa_i)-E \right]=0.
\end{equation}
For each value of the energy we have  $\kappa_i$ $(i=1,..,4l)$ solutions, 
and the generic eigenfunction
is:
\begin{equation}
\psi_{so}(x,y;E)=\sum_{i=1}^{4l} b^{so}_i e^{i \kappa_i x}\sum_{m \sigma} c^{(i)}_{m, \sigma} \chi_{m}(y)\phi_\sigma
\label{psiSO}
\end{equation}
where the coefficients $c^{(i)}_{m,\sigma}$ have to be determined numerically (the coefficients $b^{so}_i$ are determined by imposing the matching conditions
listed below). We notice, 
with reference to Fig.~\ref{nw-band-structure}, that the number of real $\{\kappa_i\}$ in 
the SO region may vary according to the value of the Fermi energy. As we will show later, changing 
the number of real  $\{\kappa_i\}$ without affecting 
the number of propagating channels in the leads,  greatly affects the conductance of the system 
when it is  in the normal state, and in 
turn the Josephson current when the leads are superconducting. 

As we consider one open transport channel in the leads with two spin orientations, 
the scattering matrix $S_e$ introduced in the previous Section is given by:  
\begin{equation}
S_e=\left (
\begin{array}{cc}
r^L & t^{RL} \\
t^{LR} & r^{R}
\end{array}
\right),
\end{equation}
with
\begin{equation}
r^L=\left (
\begin{array}{cc}
r^L_{1\uparrow,1\uparrow} & r^L_{1\uparrow,1\downarrow} \\
r^L_{1\downarrow,1\uparrow} &r^L_{1\downarrow,1\downarrow} 
\end{array}
\right),
\end{equation}
and similarly for $r^{R}, t^{RL}$ and $ t^{LR}$.

In order to obtain the total $S$-matrix, one has to compute  all the reflection and transmission 
coefficients, by matching the wave function in Eq.~\ref{psiSO} with the one in the leads 
for any possible choice of scattering boundary conditions.  For instance, 
let us consider explicitly  the case of a spin-up  particle incoming from the left-hand side. 
In this case the wave functions within L and R are respectively given by:
\begin{multline}
\psi_L(x,y;E)=e^{i k_1 x} \chi_{1}(y)\phi_\uparrow+r^{L}_{1 \uparrow,1 \uparrow}e^{-i k_1 x}\chi_{1}(y)\phi_\uparrow \\
+ r^{L}_{1 \downarrow,1 \uparrow}e^{-i k_1 x}\chi_{1}(y)\phi_\downarrow+\sum_{\sigma=\uparrow, 
\downarrow;i=2,..,n} d^{L}_{i,\sigma} \,e^{k_i x}\chi_{i}(y)\phi_\sigma
\end{multline} 
\begin{multline}
\psi_R(x,y;E)=t^{RL}_{1 \uparrow,1 \uparrow}e^{i k_1 x} \chi_{1}(y)
\phi_\uparrow+t^{RL}_{1 \downarrow,1 \uparrow}e^{i k_1 x}\chi_{1}(y)\phi_\downarrow \\
+\sum_{\sigma=\uparrow, \downarrow;i=2,..,n} d^{R}_{i,\sigma} \,e^{-k_i x}\chi_{i}(y)\phi_\sigma
\end{multline} 
with  $k_{1}=[2m(E-\hbar\omega/2)]^{1/2}/\hbar$ 
and $k_{i}=[2m(\hbar\omega(i+1/2)-E)]^{1/2}/\hbar$ for $\{i=2,...,l\}$.

The wave function at  $x=\pm \mathcal{L}$  must be continuos, its derivative with 
respect to $x$  must be, in general,    discontinuous to account the non perfect  transparency at the 
interfaces and the SO interaction (cfr. Eq.~\ref{hamiltonian}). 
Projecting the  equations corresponding to the  matching conditions onto the basis states $\chi_m(y)\phi_\sigma$
($m=1,...,l;\sigma=\uparrow,\downarrow$) we obtain the following set of equations:

\begin{equation}
\int_{-\infty}^{+\infty}\chi_m^*(y)\phi_\sigma^{\dag}\left[\psi_L(-\mathcal{L},y)-\psi_{so}(-\mathcal{L},y)\right]dy=0,
\end{equation}
\begin{equation}
\int_{-\infty}^{+\infty}\chi_m^*(y)\phi_\sigma^{\dag}\left[\psi_R(\mathcal{L},y)-\psi_{so}(\mathcal{L},y)\right]dy=0.
\end{equation}
\begin{multline}
\int_{-\infty}^{+\infty}\chi_m^*(y)\phi_\sigma^{\dag}\Big\{\partial_x \psi_{so}(-\mathcal{L},y)-\partial_x 
\psi_{L}(-\mathcal{L},y) \\ -\left[
\frac{i m}{\hbar^2}(\alpha \sigma_y-\beta \sigma_x)+\frac{2m \delta_a}{\hbar^2} \right]\psi_{so}(-\mathcal{L},y) \Big\}dy=0,
\end{multline}
\begin{multline}
\int_{-\infty}^{+\infty}\chi_m^*(y)\phi_\sigma^{\dag}\Big\{\partial_x \psi_R(\mathcal{L},y)-\partial_x \psi_{so}(\mathcal{L},y) \\ +\left[
\frac{i m}{\hbar^2}(\alpha \sigma_y-\beta \sigma_x)-\frac{2m \delta_b}{\hbar^2} \right]\psi_{so}(\mathcal{L},y) \Big\}dy=0.
\end{multline}
Therefore, we have a set of $8l$ equations which we solve numerically to determine the corresponding $S$ matrix
elements elements. Repeating  the calculation for each possible incoming channel we construct the 
complete scattering matrix $S_e$ as function of the energy $E$ which we set to $E_F$ in the following as we are interested only in on-shell scattering matrices.

\section{results and discussion}

In this section we use the scattering matrix of the normal region   to derive the
Andreev spectrum using Eq.~\ref{determinant}, from which we eventually derive  
the Josephson current as function of the phase difference between the two superconductors. 
We study the Josephson current at $\varphi=0$, i.e. the anomalous Josephson current $I_a$, as a
function of the Fermi energy.
To gain more information,  we also show the  normal conductance  $G=G_{0}\mbox{Tr}~t^\dag t$ (with $G_0 = e^2/h$) 
of the corresponding normal system in the same range of values of the Fermi energy.
As we show in details below, the AJE is accompanied by a change of the normal 
conductance  in correspondence of the opening of a new transport channel 
in the nanowire region.

\begin{figure}[h]
\includegraphics[width=1.1\linewidth]{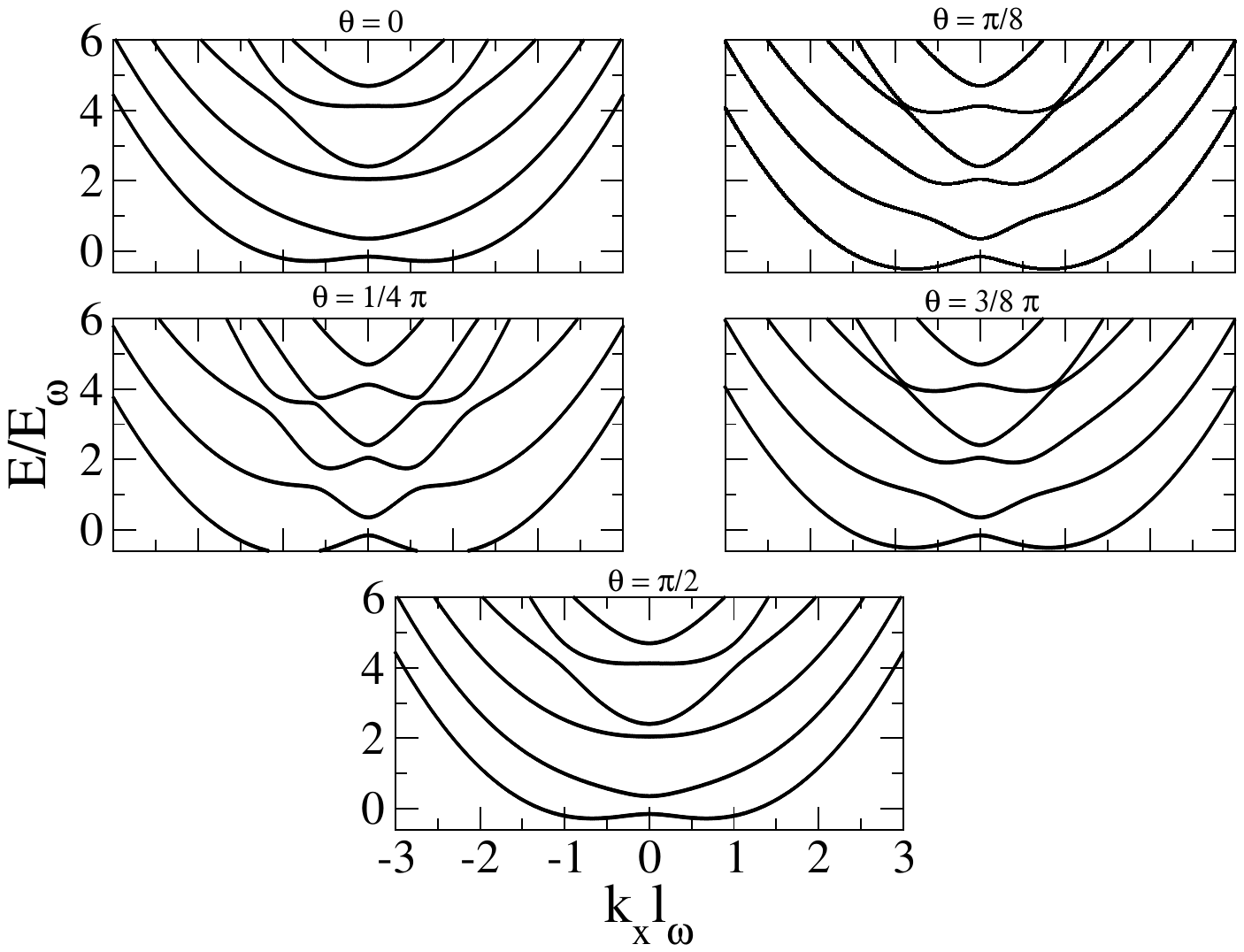}
\caption{Here we show the band structure of the nanowire for several values of the angle $\theta$. We take
the strength of the SO interaction $m \zeta l_\omega/\hbar^2=0.9$ and 
the transverse magnetic field $g \mu_B B/E_\omega=0.6$.}
\label{bandstructure}
\end{figure}

\begin{figure}[h]
\includegraphics[width=0.9\linewidth]{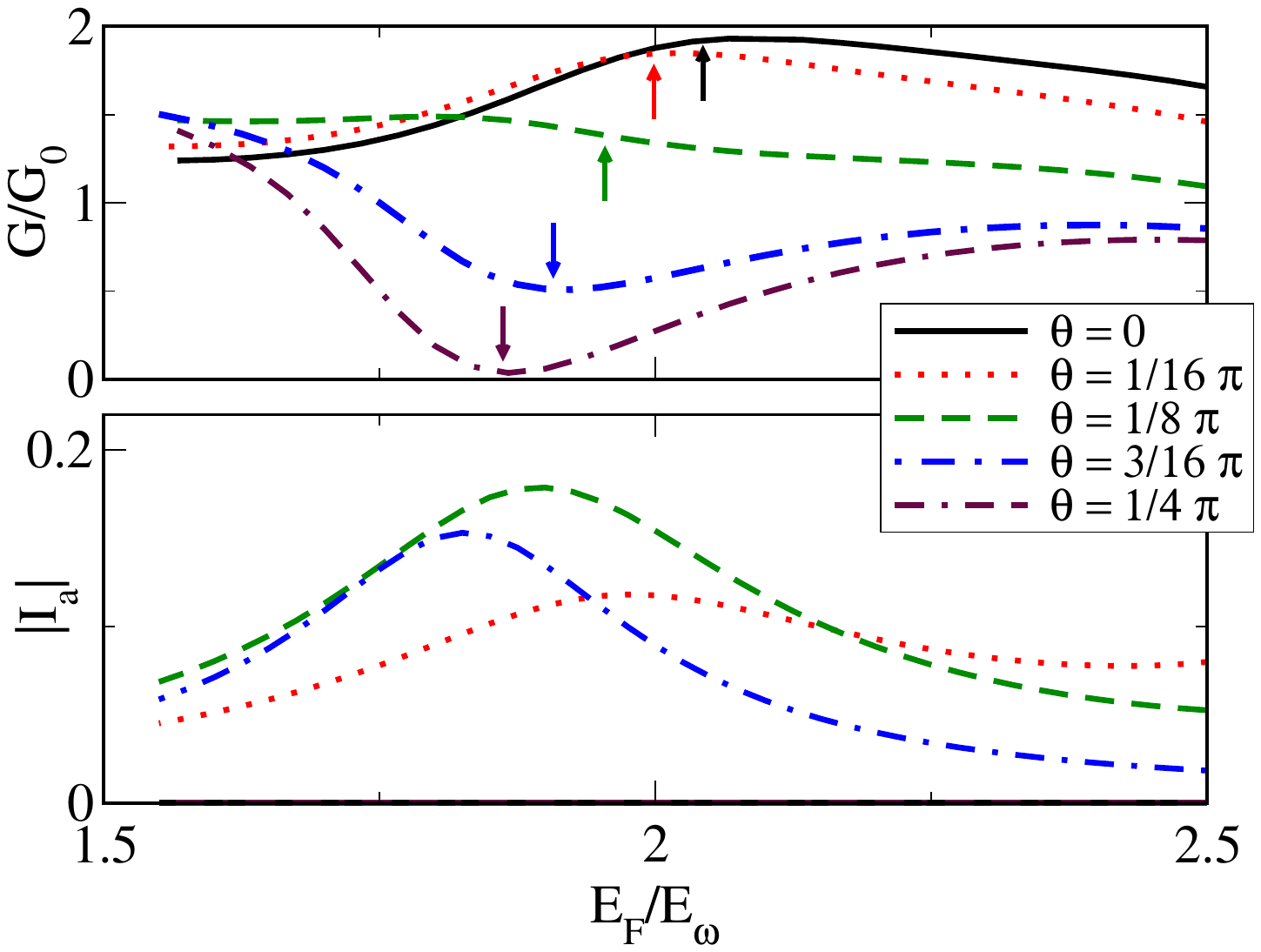} 
\caption{In panel (a) we show the total normal conductance in absence of superconductivity as 
function of the chemical potential for several values of $\theta$. The arrows indicate the position of the bottom of the band in the nanowire region.
In panel (b) we show the Josephson current at $\varphi=0$ also as function of the chemical potential 
for several values of $\theta$. Notice that for $\theta=0,\pi/4,\pi/2$ the Josephson current at $\varphi=0$ is 
always zero (see main text). The hamiltonian parameters are the same as in Fig. \ref{nw-band-structure}. The length of the junction is  $\mathcal{L}=2.3l_\omega$, and the strength of the delta-barriers are $2 m  l_\omega \delta_a/\hbar^2 =0.2$, $2 m \omega l_\omega \delta_b/\hbar^2 =-0.8$ }
\label{anomalous}
\end{figure}

In Fig.\eqref{bandstructure} the band structure is reported, with increasing ratio between the strength $\alpha $ of the Rashba  and the strength  $\beta $ of the Dresselhaus term. We parametrize them as $\alpha=\zeta \cos (\theta)$ and $\beta=\zeta \sin(\theta)$. Energies are plotted in units of $E_\omega = \hbar \omega /2 $. 
We notice that the Hamiltonian (\ref{hamiltonian}) satisfies the relation $U  H(\theta) U^{-1}=H(\pi/2-\theta)$ with $U=\exp\{i \pi  \sigma_z/4\} K$,
this implies that the normal conductance satisfies $G(\theta)=G(\pi/2-\theta)$, and explains the band structure illustrated in Fig. \ref{bandstructure}.
 In Fig.\ref{anomalous} in the top and bottom panel we plot  the normal conductance and the anomalous Josephson current, respectively, for different values of the angle $\theta$ from 0 to $\pi/4$.  
The operator $U$ acting on  the Bogoliubov-de Gennes Hamiltonian induces the relation $U\mathcal{H}_{BdG}(\theta,\varphi)U^{-1}=\mathcal{H}_{BdG}(\pi/2-\theta,-\varphi)$ thus  the Josephson current must obey the following constraint  $I(\theta,\varphi)=-I(\pi/2-\theta,-\varphi)$.  We have $I_a(\theta)=-I_a(\pi/2-\theta)$, in particular the anomalous Josephson current is zero for $\theta=\pi/4$. Besides for $\theta=\pi/4$ we find that the anomalous Josephson current is also zero for $\theta=0$ (pure Rashba) and $\theta=\pi$ (pure Dresselhaus). 
 
For $\theta=0$ the above statement can be understood as follows: even in the presence of delta barriers which breaks inversion symmetry in the $x$ direction the  operator $O_1=R_y K$ acting on the Bogoliubov-de Gennes Hamiltonian 
gives $O_1\mathcal{H}_{BdG}(\varphi)O_1^{-1}=\mathcal{H}(-\varphi)$, hence $I_a=0$.  
For the $\theta=\pi/2$ case, the operator  $O_2=R_y K \sigma_z$ acting 
on the Bogoliubov-de Gennes Hamiltonian gives $O_2\mathcal{H}_{BdG}(\varphi)O_2^{-1}=\mathcal{H}(-\varphi)$, and again $I_a=0$.

\begin{figure}[h]
\includegraphics[width=1\linewidth]{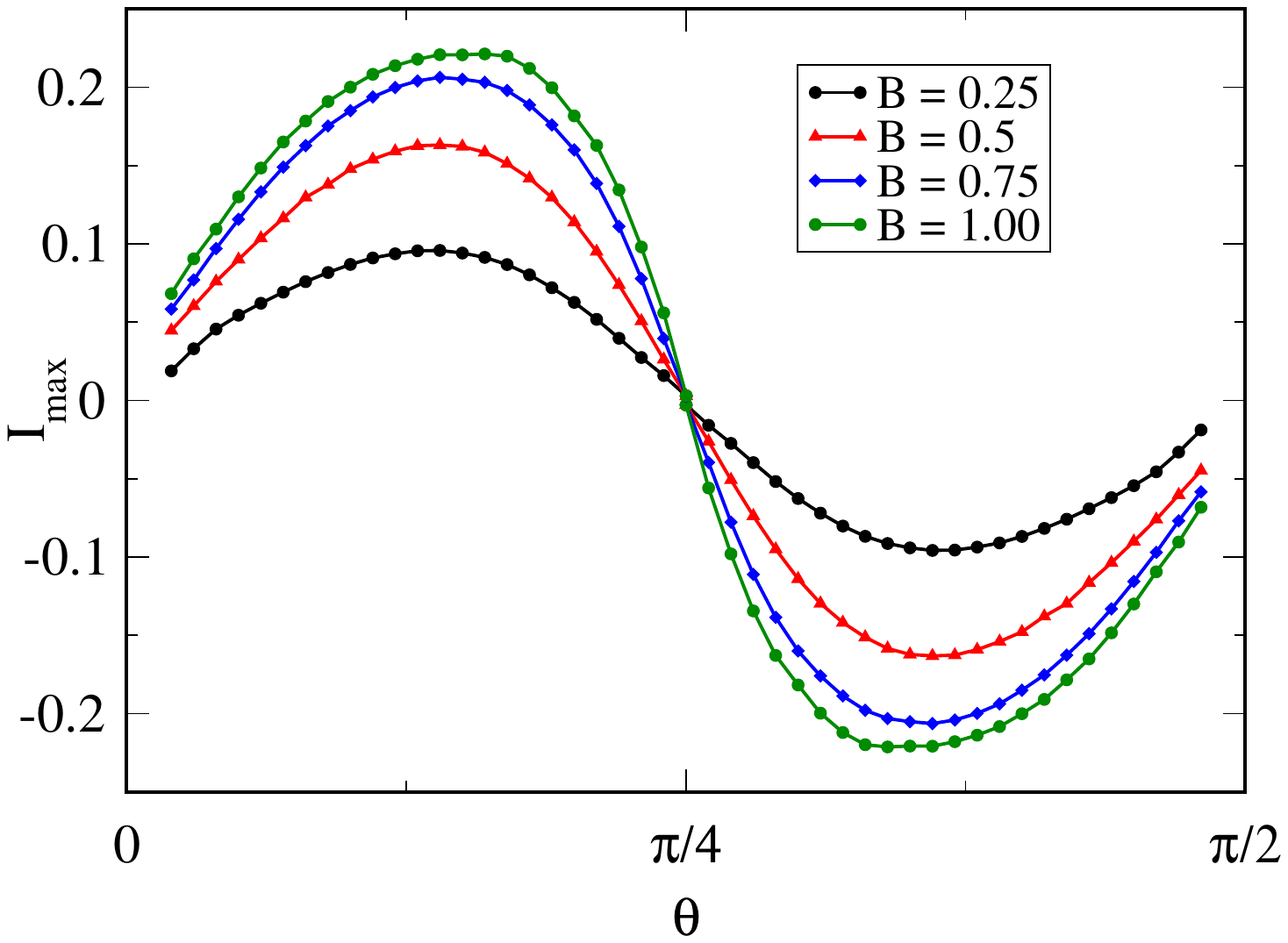}
\caption{In this figure we show the maximum anomalous Josephson current $I_{max}$ (cfr. main text)
obtained spanning the energy range $E_F=1.5E_\omega$ to $E_F=2.5E_\omega$ as function of the angle $\theta$
for several values of the magnetic field in units of $E_\omega/g \mu_B$.}
\label{Imax-vs-theta}
\end{figure}

For intermediate values of $\theta$ ($\theta\ne \{0,\pi/4,\pi/2\}$)  we do find 
anomalous Josephson effect. Let us consider as an example $\theta=\pi/8$ (cfr. Fig. \ref{anomalous}). 
As the chemical potential makes the band $c$ accessible the conductance decreases but, most importantly,  
the anomalous Josephson current is maximum. The latter effect is independent of the particular choice of $\theta$.

To characterize the dependance of the anomalous Josephson current on the magnetic field and on the angle 
$\theta$, we introduce the quantity  $I_{max}$ as the maximum anomalous current $I_a$ in the interval from $E_F=1.5~E_{\omega}$ 
to $E_F=2.5~E_{\omega}$ (see Fig.~\ref{Imax-vs-theta}).
Such a range is sufficient to explore the opening of the band $c$ for all the values of $\theta$ and $B$ considered here.
We find that for a fixed value of $\theta$ the absolute value of $I_{max}$ is enhanced by the magnetic field and that 
for a fixed value of the 
magnetic field the largest value of $I_{max}$ (in absolute value) is found for $\theta\simeq \pi/8$.

\begin{figure}
\includegraphics[width=1\linewidth]{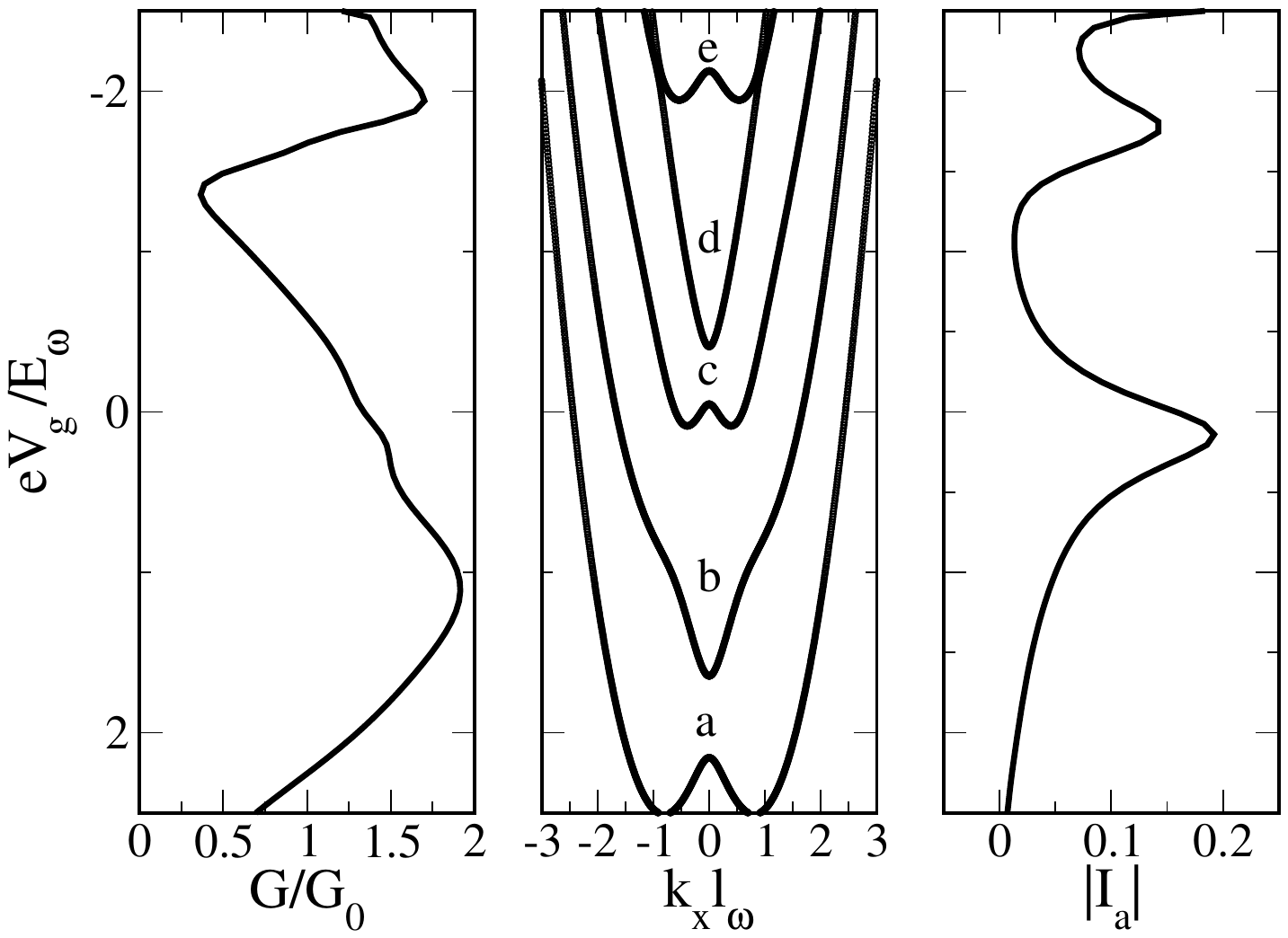}
\caption{In this figure the Fermi energy is set to $E_F=2E_\omega$ and we consider a rigid shift
of the bands of the nanowire region by applying a gate potential $V_g$. $\theta=\pi/4$ and the remaining parameters are 
the same as in Fig.~\ref{nw-band-structure}.
We plot the normal conductance and the Josephson current at  $\varphi=0$.}
\label{gate_sweep}
\end{figure}

Our study is limited to the case $N=1$ so that only one band (twice spin degenerate) is open in the leads. In order to address the effect of
the nanowire's higher bands, keeping the Fermi energy constant, we introduce a gate in the normal region.
 We modify the Hamiltonian of Eq.~\ref{hamiltonian} as $H\rightarrow H+e V_g$. 
In Fig.~\ref{gate_sweep} we set the value of the Fermi energy to $E_F=2E_\omega$ and study the normal conductance and the anomalous 
Josephson current as function of the gate potential $V_g$. We find that the anomalous Josephson current exhibits a similar behaviour as 
found before for the case of band $c$ (cfr Fig. \ref{anomalous}), indeed it is maximum when the bottom of band  $e$  is energetically accessible (cfr. Fig.~\ref{gate_sweep}).
One might speculate that no anomalous Josephson is found at the opening of band $b$ and $d$ because both spin polarisations
are fully contributing to the transport, alas, such a simple picture is not correct as either spin polarisation (and helicity) 
are not good quantum number in the nanowire region.
Notice that this approach, albeit instructive, is not equivalent to changing the Fermi energy in the leads.

As a last remark we would like to stress
that in our setup the magnetic field is perpendicular to the plane containing the nanowire. If one would change 
the  orientation of the magnetic
field respect to the nanowire plane one may find the angle $\theta$ which maximizes the anomalous Josephson 
effect to be different from $\pi/4$. Notice that with an in-plane magnetic field (not considered here) we should observe AJE even with 
pure Rashba or pure Dresselhaus spin orbit interaction.

\section{conclusions}
We have studied the DC Josephson effect in a superconductor/nanowire/superconductor junction. 
The nanowire has a strong spin orbit interaction of Rashba and Dresselhaus type, moreover  a
perpendicular Zeeman field is applied.
We include a finite number of channels in the calculation.
Although our approach can be extended to a generic number $N$ of open channels in the leads,  we have discussed in details the case of a $N=1$, but allowing the number of open channels in the normal region region to be larger than 1. We have found that a necessary condition for 
the anomalous Josephson effect (in our specific setup: i.e.  with a perpendicular magnetic field) is the simultaneous presence of Rashba 
and Dresselhaus spin orbit interaction together with an asymmetric choice of the barriers at the 
interfaces between normal and superconductiong regions. This condition may seem to be restrictive but, however, it is completely satisfied in actual experiments.
Indeed, for a small number of impurities in the system we expect these to mimimc the role of the delta potentials in our model, hence giving rise to the AJE. On the other hand as this number increases we should move towards a self-averaging regime where the reflection symmetry is restored and the AJE suppressed.

Most important for our results, we observe that, because of the spin orbit interaction the electronic band
structure of the nanowire region is lowered with respect to the one in the leads, by carefully tuning the Fermi
level, such to remain in the $N=1$ case, we find that  the anomalous Josephson effect is maximum in correspondence
of the opening of a new band in the nanowire region. We do expect this phenomenology to be found also in the case 
$N>1$; this will be object of further investigation.   
It is important to stress that the condition found here for the anomalous Josephson effect, namely the opening of a
new band in the normal region, is different to that found for instance in Refs.[\onlinecite{Krive:2004,Yokoyama:2014}],
where it can be ascribed  to the asymmetry of the 1D electronic spectrum due to the coupling of transverse bands by 
the spin orbit interaction.  

\section*{Acknowledgements}
We acknowledge enlightening discussions with P.W. Brouwer, V. Marigliano Ramaglia, Yu. V. Nazarov, F. Trani and T. Yokoyama.
Financial support from FIRB 2012 Projects "HybridNanoDev" (Grant No. RBFR1236VV)  is gratefully acknowledged.


\begin{thebibliography}{52}
\bibitem[{not({\natexlab{a}})}]{note}
\bibinfo{note}{The same system can be addressed with pure Rashba SO or
  Dresselhaus SO, but allowing for a magnetic field with an in-plane
  component.}
\end{thebibliography}

\end{document}